\documentclass{elsarticle}
\usepackage{graphicx}
\usepackage{bm} 
\usepackage{amssymb}
\begin{document}
\begin{frontmatter}
\title {A CsI(Tl) detector array for the measurement of light charged particles in heavy-ion reactions}
\author[a]{P. C. Rout }
\author[b]{V. M. Datar}
\author[d]{D. R. Chakrabarty}
\author[d]{Suresh Kumar}
\author[a]{E. T. Mirgule}
\author[a]{A. Mitra}
\author[c]{V. Nanal} 
\author[a]{R. Kujur}
\address[a]{Nuclear Physics Division, Bhabha Atomic Research Centre, Mumbai-400085, India}
\address[b]{INO~Cell, Tata Institute of Fundamental Research, Mumbai-400005, India}
\address[d]{Ex-Nuclear Physics Division, Bhabha Atomic Research Centre, Mumbai-400085, India}
\address[c]{Department of Nuclear and Atomic Physics, Tata Institute of Fundamental Research, Mumbai-400005, India}

\begin{abstract}
An array of eight CsI(Tl) detectors has been set up to measure the light charged particles in nuclear reactions using heavy ions from the Pelletron Linac Facility, Mumbai. The energy response of CsI(Tl) detector to $\alpha$-particles from 5 to 40~MeV is measured using radioactive sources and the $^{12}$C($^{12}$C,~$\alpha$) reaction populating discrete states in $^{20}$Ne. The energy non-linearity and the count rate effect on the pulse shape discrimination property have also been measured and observed the deterioration of pulse shape discrimination with higher count rate.

\end{abstract}
\begin{keyword}
CsI(Tl) scintillator,  pulse shape discrimination, $\alpha$ spectra in 
$^{12}$C+$^{12}$C and $^{7}$Li+$^{197}$Au reactions 
\end{keyword}
\end{frontmatter}
\section{Introduction}
CsI(Tl) scintillation crystals have been widely used for both $\gamma$-ray and charged particle spectroscopy. These scintillators have higher density and constitute elements of higher atomic number than NaI(Tl) scintillation crystals. The scintillation light of the CsI(Tl) detector is composed of two components, a fast component with decay time of $\sim$0.6$~\mu$s and a slower one with decay time of $\sim$3.5~$\mu$s~\cite{knoll,sg}. The amplitude ratio of these components as well as the decay time constant of the fast component vary with the type of radiation interacting with the crystal. These properties of CsI(Tl) can be used to identify the particle that caused the scintillation. Many techniques such as the charge comparison~\cite{jal}, zero cross over time~(ZCT)~\cite{rfu} and ballistic deficit~\cite{gal} are used for the charged particles identification. In addition to the above techniques, The FPGA based digital signal processing has also been employed for the particle identification using pulse shape analysis~\cite{skulski}. In the charge comparison method, the ratio of integrated charge of linear signal for a short time~(few hundreds of ns) to long time~(few thousands of ns) window provides the identification of charged particles. The amplitude degradation at the output of a pulse shaping circuit due to the finite rise time of the input pulse is used for the particle identification in the ballistic deficit method.  The zero crossover time of the amplified bipolar signal with respect to the fast timing is used for the particle identification in the ZCT technique.
%
\begin{figure}
\begin{center}
\includegraphics[scale=0.5]{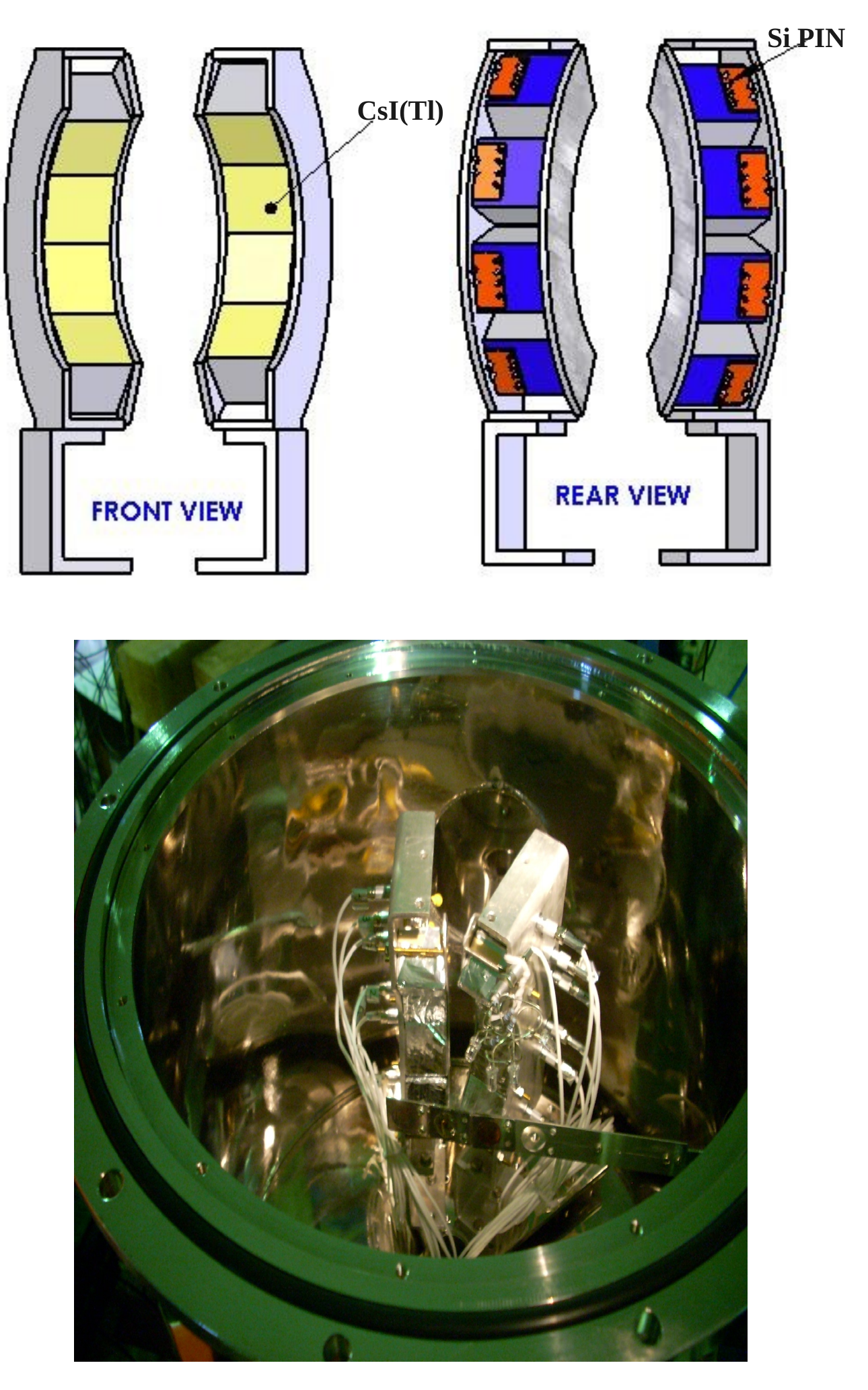}
\caption{A schematic of the CsI(Tl) detector array~(upper) and a photograph of the detectors mounted in a thin wall multi-purpose chamber~(lower)}
\label{fig01asm}
\end{center}
\end{figure}
In this paper, we report the characterization of the CsI(Tl) detector. The paper is organized mainly in four sections. The second section describes details of the detector array. In the next section the measured alpha particles in $^{12}$C($^{12}$C,~$\alpha$)$^{20}$Ne reaction at E($^{12}$C)~=~24,30 and 40~MeV are presented. The subsequent section explains the count rate effect on the PSD in the $^{7}$Li+$^{197}$Au at E($^{7}$Li)~=~30~MeV  which is followed by a summary. 
\section{Description of CsI(Tl) Detector Array}
%
\begin{figure}
\begin{center}
\includegraphics[scale=.60]{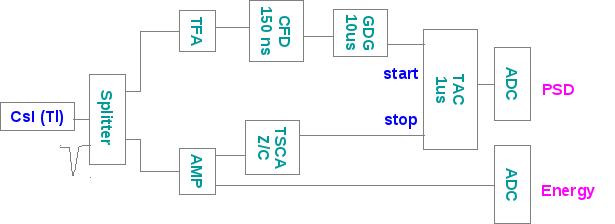}
\caption{Basic electronic block diagram for the PSD using laboratory made Zero Cross over module(Hex-ZCT). The electronic modules includes, TFA:~Time filtered amplifier, AMP:~ Spectroscopic amplifier, CFD:~ Constant fraction discrimination, TSCA:~Timing single channel analiser/Hex-ZCT, TAC:~Time to amplitude converter, GDG:~Gate and delay generator and ADC:~Analogue to digital converter.}
\label{zct_bd}
\end{center}
\end{figure}
An array of eight CsI(Tl) scintillators, each coupled to a Si(PIN) photodiode was assembled to detect the charged particles in the heavy ion induced reaction at the Pelletron Linac Facility, Mumbai. Since the size of the photodiode is small a compact closed packed configuration can be made. The detectors were grouped in to two arrays, each consisting of four detectors and mounted in aluminum frames as shown in Fig.~\ref{fig01asm}. The CsI(Tl) detector  array will be used for the measurement of charged particle by pulse shape discrimination in the reaction with weakly bound nuclei and also useful for the coincidence measurements involving neutrons and gamma rays~\cite{pcr-prl,pcr-nim}.
The CsI(Tl) scintillators coupled to Si(PIN) photodiode have been procured from M/s SCIONIX, Holland~\cite{scionix}. 
The active area of the CsI(Tl) is 25$\times$25~mm$^2$ and  the thickness is 10~mm. The coupling of Si-PIN photodiodes~(Hamamatsu S3204-08) to the scintillators was done using a perspex light guide of dimension 25$\times$25$\times$15~mm$^3$. The sensitive area of the photodiode is 18$\times$18~mm$^2$. Some of its important features are good energy resolution, good stability, fast response, low capacitance and high quantum efficiency (85\% at peak wavelength 540~nm). The signal readout was taken through a high gain~($\sim$9~V/pC) charge sensitive preamplifier  mounted close to the Si-PIN photodiode for further processing. The preamplifier is vacuum compatible and has low power dissipation ($\sim$100~mW). The DC offset in the preamplifier signal output was eliminated by using a capacitor ($\sim$6~$\mu$F) and a battery was used to supply 12~V to preamplifier in order to reduce the pick-up problem and, thus, to improve the signal to noise ratio. 
%
\begin{figure}
\begin{center}
\includegraphics[scale=0.7]{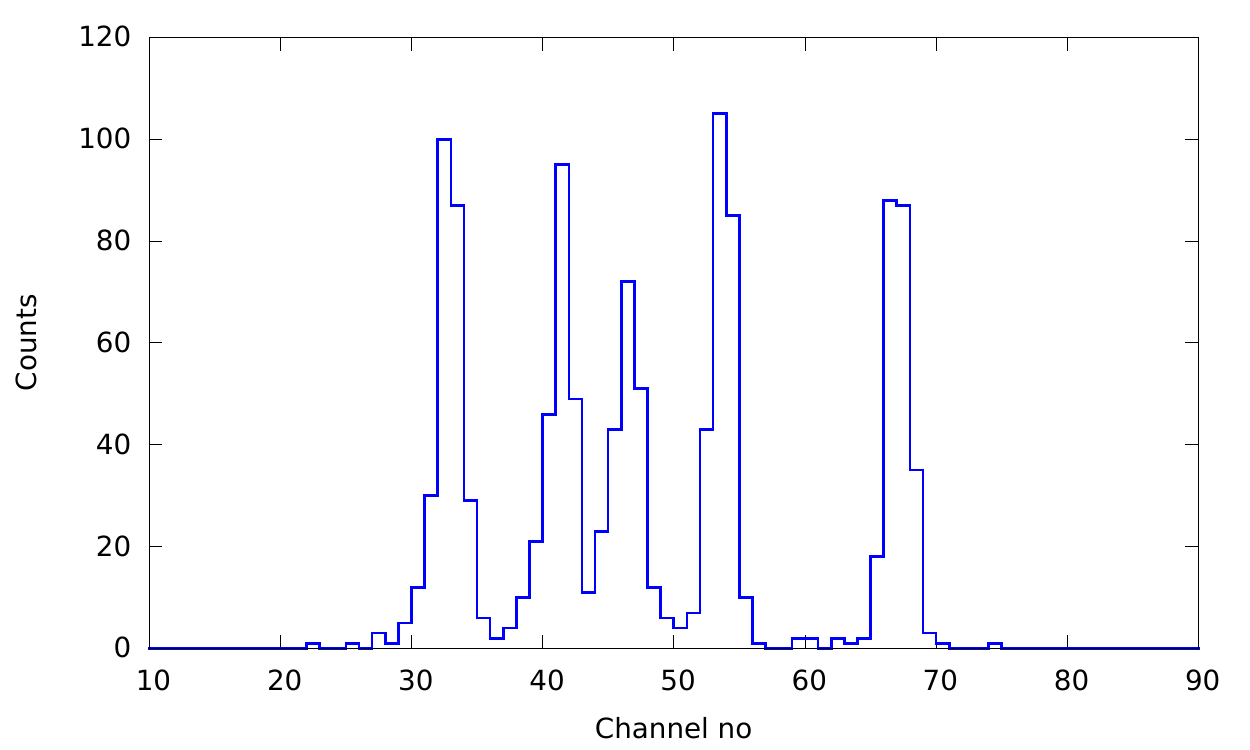}
\caption{Typical alpha spectrum of the CsI detector using $^{229}$Th source. The peaks corresponds to the alpha particles with energies 4.9, 5.8, 6.3, 7.1 and 8.4 MeV.}
\label{en_Th} 
\end{center}
\end{figure}
A six channels ZCT (Hex-ZCT) module was specially designed and fabricated for the CsI(Tl) detectors to identify particles by ZCT technique. The measured figure of merit~(M), which is defined as the ratio of peak separation between two particles to the sum of their full width at half maxima(FWHM), observed to be comparable with that obtained using a commercial module~\cite{kujur}. 

The typical block diagram used for the measurement of the PSD and energy of the particles is shown in Fig.~\ref{zct_bd}. The preamplifier signal was split, one fed to the constant fraction discriminator through the fast amplifier for the start time of Time to amplitude converter(TAC) and second part fed to the amplifier for the measurement of energy and zero cross over time to stop of the TAC for pulse shape discrimination. The measured alpha spectrum for the $^{229}$Th source is shown in Fig.~\ref{en_Th}. Typical energy resolution of these detectors is about 6\% at $\sim$5~MeV for alpha particles.
\begin{figure}
\begin{center}
\includegraphics[scale=0.45]{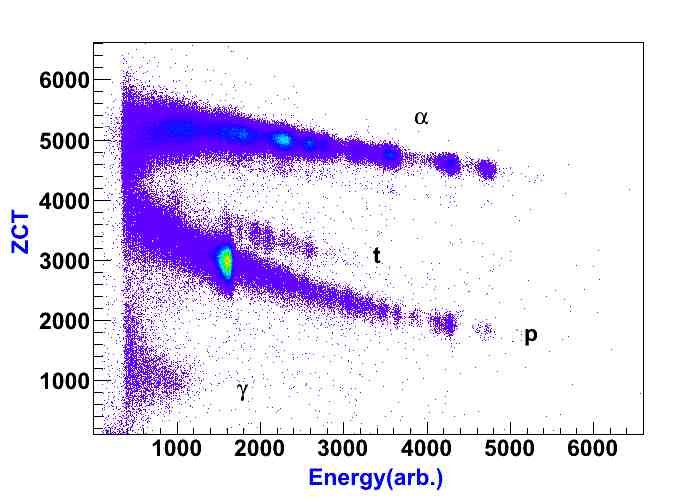}
\caption{A 2D spectrum between the ZCT and the energy of CsI(Tl) detector which shows the particles emitted in the $^{12}$C+$^{12}$C reaction are well separated by the pulse shape discrimination technique.}
\label{PSD12C}
\end{center}
\end{figure}
%
\begin{figure}
\begin{center}
\includegraphics[scale=0.55]{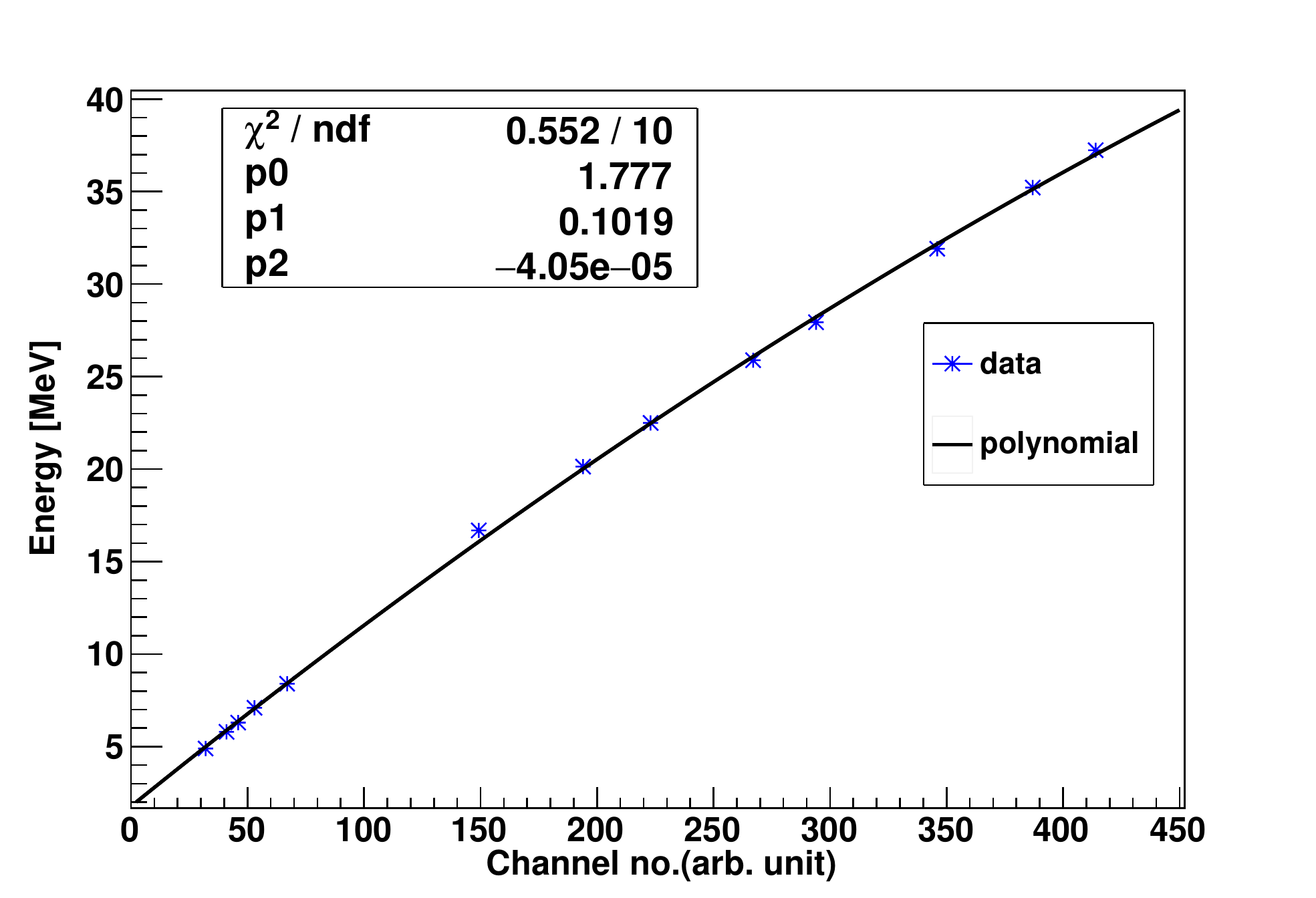}
\caption{Measured calibration curve for alpha particles up to $\sim$ 40~MeV.}
\label{calib_alpha}
\end{center}
\end{figure}
\section{Response of CsI(Tl) detector to $\alpha$-particles}
The amount of light produced by a CsI(Tl) scintillators depends on the energy, charge and mass of the particle interacting with the scintillator. The light output of the scintillator is related to the stopping power (energy loss per unit length) of the particle in the crystal and is described by the Birks relation~\cite{knoll}. It is a non-linear function of the energy of the alpha particles. Thus, in order to use these detectors for the alpha spectroscopy, it is necessary to measure the light output over a wide energy range for alpha particles. The non-linearity of the light output of  CsI(Tl) detector in the energy region from $\sim$5  to 40 MeV has been  measured for $\alpha$-particles. The alpha response from $\sim$5 to 8~MeV was measured using $^{229}$Th source and  is shown in Fig.~\ref{en_Th}. The response at higher energies was measured with alpha particles from $^{12}$C($^{12}$C,~$\alpha$)$^{20}$Ne reaction  populating discrete states in $^{20}$Ne.

%
\begin{figure}
\begin{center}
\includegraphics[scale=0.6]{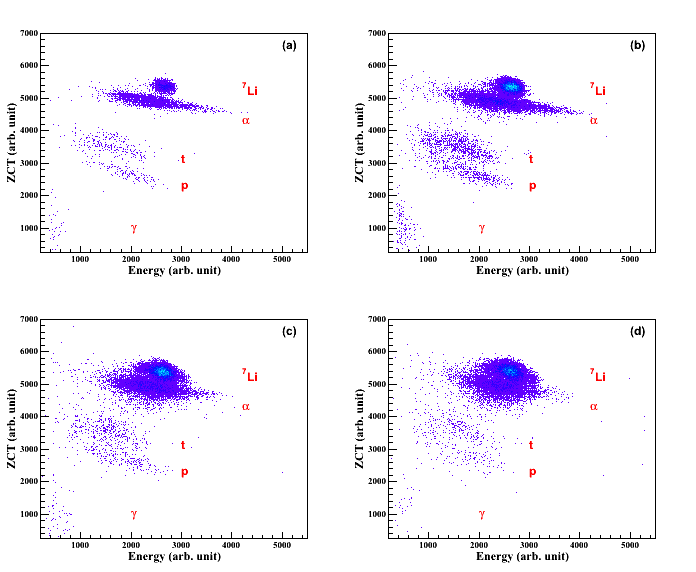}
\caption{Pulse shape discrimination spectra of CsI(Tl) detector for $^{7}$Li+$^{197}$Au reaction with increasing count rates:(a)~1.6~kHz, (b)~3.5~kHz, (c)~6.7~kHz and (d)~9.2~kHz.}
\label{psd02}
\end{center}
\end{figure}
%
\begin{figure}
\begin{center}
\includegraphics[scale=3.0]{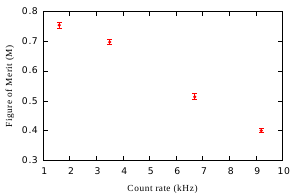}
\caption{Figure of merit as function of the count rate}
\label{fom2}
\end{center}
\end{figure}
The experiment was carried out to measure alpha particles in ($^{12}$C,~$\alpha$) reaction at 24, 30 and 40~MeV $^{12}$C beam using the Mumbai PLF . The carbon target backed by 1 mil thick Ta foils was mounted 9.5~cm upstream of the centre of the reaction chamber  and the detectors were placed at 4.5~cm from the centre. The detectors were brought to 0$^\circ$ to reduce the kinematic energy spread and alpha particles were detected by the ZCT technique. A 2D spectrum between ZCT and energy of the particles for this reaction is shown in Fig.~\ref{PSD12C}. The gamma rays, proton, triton and alpha particles were clearly identified. The projected alpha energy spectra were corrected for the energy loss in the Ta foil using the energy loss of alpha particles in Ta calculated using SRIM~\cite{srim}. The measured energy  calibration for alpha particles is shown in Fig.~\ref{calib_alpha}. It can be seen from the figure that the calibration is not linear over the full energy range and can be fit to a polynomial function of light output~\cite{ylar}. The measured non-linearity in energy of alpha particles has been used to calibrate the CsI(Tl) detectors in the alpha coincidence experiment.

\section{Effect of count rate on Pulse shape discrimination}
An experiment was performed to study the count rate effect on the PSD in the CsI(Tl)  detectors at the Mumbai Pelletron Linac Facility.  The beam of $^7$Li at 30~MeV from the Pelletron bombarded a self-supported $^{197}$Au target and the beam current was varied from 4enA-24enA. The parameters, energy and ZCT were recorded in an event-by event mode with LAMPS data acquisition system~\cite{lamps}. The FiG~\ref{psd02}(a) shows the particles from proton to $^7$Li are well separated in a 2D figure between energy and ZCT. The discrimination between alpha and $^7$Li become worse with increasing count rate as shown in Fig~\ref{psd02}(a)-(d). Quantitaively, the extracted figure of merit~(M) obtained from the projected ZCT spectrum decreases with increasing count rate. Fig~\ref{fom2} shows the effect of count reate on the figure of merit. This could be due to the fact that zero cross over time deteriorates with high count rates due to the pile up events.

\section{Summary}
An array of eight CsI(Tl) scintillators coupled to Si-PIN diodes with preamplifier is setup to measure the light charged particle in the nuclear reaction  using  heavy ions from the PLF, Mumbai. The nonlinear alpha response of the detector using Th source and $^{12}$C($^{12}$C,~$\alpha$) reaction up to 40~MeV is measured. The pulse shape discrimination by ZCT method is used for the particle identification in the $^7$Li induced reaction on heavy target and observed that the PSD deteriorates with increasing count rates beyond $\sim$3~kHz. 

\noindent {\bf Acknowledgements}\\
We thank the Pelletron staff for the smooth operation of the accelerator and  R. Parui for his help during the experiment.

\end{document}